\documentclass{article}

\usepackage{arxiv}

\usepackage[utf8]{inputenc} 
\usepackage[T1]{fontenc}    
\usepackage[hidelinks]{hyperref}       
\usepackage{url}            
\usepackage{booktabs}       
\usepackage{amsfonts}       
\usepackage{nicefrac}       
\usepackage{microtype}      
\usepackage{lipsum}		
\usepackage{graphicx}
\usepackage[style=phys,sorting=none,backend=biber,doi=true,isbn=false,url=false,firstinits=true,maxcitenames=2,minbibnames=3,natbib=true]{biblatex}
\usepackage{doi}

\usepackage{blindtext}
\usepackage{amsmath}

\usepackage{siunitx}
\usepackage{graphicx}
\usepackage{nameref}
\usepackage{subcaption}
\usepackage{titlesec}
\usepackage{glossaries}
\usepackage{cleveref}
\usepackage{float}
\crefname{figure}{Fig.}{Figs.}
\Crefname{figure}{Figure}{Figures}
\crefname{section}{Sec.}{Secs.}
\Crefname{section}{Section}{Sections}
\crefname{table}{Table}{Tables}
\Crefname{table}{Table}{Tables}
\crefname{equation}{Eq.}{Eqs.}
\Crefname{equation}{Equation}{Equations}
\crefname{appendix}{Appendix}{Appendices}
\Crefname{appendix}{Appendix}{Appendices}
\crefname{algorithm}{Algorithm}{Algorithms}
\Crefname{algorithm}{Algorithm}{Algorithms}
\crefname{listing}{Listing}{Listings}
\Crefname{listing}{Listing}{Listings}

\makeglossaries 
\newglossaryentry{computer}{
	name=computer,
	description={is a programmable machine that receives input, stores and manipulates data, and provides output in a useful format}
}

\newacronym[longplural={Frames per Second}]{fpsLabel}{FPS}{Frame per Second}
\newacronym[longplural={Tables of Contents}]{tocLabel}{TOC}{Table of Contents}
\newacronym[longplural={Autoregressive Integrated Moving Average}]{arima}{ARIMA}{Autoregressive Integrated Moving Average}
\newacronym[longplural={Autoregressive Moving Average}]{arma}{ARMA}{Autoregressive Moving Average}
\newacronym[longplural={Autoregressive}]{ar}{AR}{Autoregressive}
\newacronym[longplural={Moving Average}]{ma}{MA}{Moving Average}
\newacronym[longplural={Probability Density Functions}]{pdf}{PDF}{Probability Density Function}
\newacronym[longplural={Cumulative Distribution Functions}]{cdf}{CDF}{Cumulative Distribution Function}
\newacronym[longplural={Probability Mass Functions}]{pmf}{PMF}{Probability Mass Function}
\newacronym[longplural={Cumulative Mass Functions}]{cmf}{CMF}{Cumulative Mass Function}
\newacronym[longplural={Autocorrelation Functions}]{acf}{ACF}{Autocorrelation Function}
\newacronym[longplural={Partial Autocorrelation Functions}]{pacf}{PACF}{Partial Autocorrelation Function}
\newacronym[longplural={High Energy Physics}]{hep}{HEP}{High Energy Physics}
\newacronym[longplural={Electromagnetic Calorimeters}]{ecal}{ECAL}{Electromagnetic Calorimeter}
\newacronym[longplural={Hadronic Calorimeters}]{hcal}{HCAL}{Hadronic Calorimeter}
\newacronym[longplural={A Toroidal LHC Apparatus}]{atlas}{ATLAS}{A Toroidal LHC Apparatus}
\newacronym[longplural={Large Hadron Collider}]{lhc}{LHC}{Large Hadron Collider}
\newacronym[longplural={Compact Muon Solenoid}]{cms}{CMS}{Compact Muon Solenoid}
\newacronym[longplural={Monolithic Active Pixel Sensors}]{maps}{MAPS}{Monolithic Active Pixel Sensor}
\newacronym[longplural=ITk Strips Data Acquisitions]{itsdaq}{ITSDAQ}{ITk Strips Data Acquisition}
\newacronym[longplural=Field Programmable Gate Arrays]{fpga}{FPGA}{Field Programmable Gate Array}
\newacronym[longplural={Depleted Monolithic Active Pixel Sensors}]{dmaps}{DMAPS}{Depleted Monolithic Active Pixel Sensor}
\newacronym[longplural={Signal-to-Noise Ratios}]{snr}{SNR}{Signal-to-Noise Ratio}
\newacronym[longplural={Digital-to-Analog Converters}]{dac}{DAC}{Digital-to-Analog Converter}
\newacronym[longplural={European Organizations for Nuclear Research}]{cern}{CERN}{European Organization for Nuclear Research}
\newacronym[longplural={Digital Electromagnetic Calorimeters}]{decal}{DECAL}{Digital Electromagnetic Calorimeter}
\newacronym[longplural={Akaike Information Criterion}]{aic}{AIC}{Akaike Information Criterion}
\newacronym[longplural={Bayesian Information Criterion}]{bic}{BIC}{Bayesian Information Criterion}
\newacronym[longplural={Comma Separated Values}]{csv}{CSV}{Comma Separated Values}
\newacronym[longplural={Maximum Likelihood Estimations}]{mle}{MLE}{Maximum Likelihood Estimation}
\newacronym[longplural={Binned Maximum Likelihood Estimations}]{bmle}{BMLE}{Binned Maximum Likelihood Estimation}
\newacronym[longplural={Mean Squared Disploacements}]{msd}{MSD}{Mean Squared Displacement}
\newacronym[longplural={Fast Fourier Transforms}]{fft}{FFT}{Fast Fourier Transform}
\newacronym[longplural={USA National Institute of Standards and Technology}]{nist}{NIST}{USA National Institute of Standards and Technology}
\newacronym[longplural={Random Number Generators}]{rng}{RNG}{Random Number Generator}
\newacronym[longplural={Pseudo-Random Number Generators}]{prng}{PRNG}{Pseudo-Random Number Generator}
\newacronym[longplural={True Random Number Generators}]{trng}{TRNG}{True Random Number Generator}
\newacronym[longplural={Electronic Random Number Generator Indicator Equipment}]{ernie}{ERNIE}{Electronic Random Number Indicator Equipment} 

\title{Time Series Analysis of DECAL Sensor Noise for the Generation of Truly Random Numbers}

\author{ 
    \href{https://orcid.org/0009-0001-3271-592X}{\includegraphics[scale=0.06]{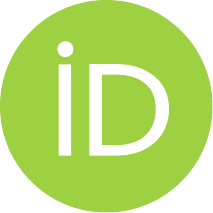}\hspace{1mm}Dimos Aslanis}\thanks{Current address: Institute for Theoretical Physics, Center for Extreme Matter and Emergent Phenomena, Utrecht University, 3584 CC Utrecht, The Netherlands.} \\
	Department of Physics\\
	National Technical University of Athens\\
	157 80, Zografou Campus, Athens, Greece\\
	\texttt{d.aslanis@uu.nl} \\
	\And
	\href{https://orcid.org/0000-0001-6080-6215}{\includegraphics[scale=0.06]{orcid.pdf}\hspace{1mm}Alex Kehagias} \\
	Department of Physics\\
	National Technical University of Athens\\
	157 80, Zografou Campus, Athens, Greece\\
	\texttt{kehagias@central.ntua.gr} \\
	\And
	\href{https://orcid.org/0000-0001-6145-7467}{\includegraphics[scale=0.06]{orcid.pdf}\hspace{1mm}Ioannis Kopsalis} \\
	Department of Physics\\
	National Technical University of Athens\\
	157 80, Zografou Campus, Athens, Greece\\
	\texttt{ikopsali@mail.ntua.gr} \\
	\And 
	Ioannis Theodonis \\
	Department of Physics\\
	National Technical University of Athens\\
	157 80, Zografou Campus, Athens, Greece\\
	\texttt{ytheod@mail.ntua.gr} \\
}

\renewcommand{\headeright}{}

\hypersetup{
pdftitle={Time Series Analysis of DECAL Sensor Noise for the Generation of Truly Random Numbers},
pdfsubject={physics, computer science},
pdfauthor={Dimos Aslanis},
pdfkeywords={Time Series Analysis, Stochastic Noise, DECAL Sensor, Truly Random Numbers},
}

\begin{document}

\maketitle

\begin{abstract}

    We explore here the stochastic behavior of the DECAL sensor's noise output, and we evaluate its potential application as a true random number generator (TRNG) using time series analysis. The main objectives are twofold: first, to characterize the intrinsic noise properties of the DECAL sensor in the absence of external stimuli, and second, to determine the feasibility of employing the sensor as a source of randomness. The collected sensor data are examined through statistical and time series methodologies, and subsequently modeled using an auto-regressive integrated moving average (ARIMA) process. This modeling approach enables the transformation of the sensor’s raw noise into a Gaussian white noise sequence, which serves as the basis for generating random bits. The resulting random numbers are subjected to a series of statistical tests for randomness, including the NIST test suite. Our findings indicate that the method produces statistically sound random numbers. However, the rate of bit generation is relatively low, limiting its practicality for real-time TRNG applications under the current configuration. Despite this limitation, the results suggest that time series modeling presents a promising framework for extracting randomness from the DECAL sensor, and that with further optimization, the sensor could serve as a reliable and effective TRNG.
\end{abstract}

\glsresetall

\keywords{Random Number Generation, Time Series Analysis, Signal Processing, DECAL Sensor}

\section{Introduction}
\label{sec:introduction}

Random number generation has played a pivotal role throughout history \cite{lecuyer_history_2017}. Modern applications include simulations \cite{landau_guide_2015,newman_monte_1999}, mathematics \cite{gentle_random_2005}, programming \cite{knuth_art_2021}, and cryptography \cite{menezes_handbook_1997}. The first dedicated \gls{rng} was developed in 1939 \cite{kendall_second_1939}, with subsequent advancements driven by the advent of electronic computing. While early random numbers were distributed via printed tables, growing demands led to the development of two main types of \glspl{rng}: \glspl{prng} and \glspl{trng}.

\glspl{prng} are deterministic algorithms widely used in simulations and programming, where reproducibility is essential. Conversely, \glspl{trng} rely on physical processes and are favored in cryptography, where reproducibility should be avoided. Outputs from both types are often post-processed to eliminate bias, with \glspl{trng} commonly serving as seeds for \glspl{prng} to enhance efficiency. Stipčević and Koç \cite{koc_true_2014} classify \glspl{trng} into four types: noise-based (e.g., Johnson or Zener noise), chaotic (e.g., lasers with deterministic initial conditions), free-running oscillators, and quantum systems (e.g., Geiger counters). The primary goal of any \gls{rng} is to produce uniformly distributed random numbers. Numbers for other distributions are typically obtained by transforming uniform distributions \cite{devroye_non-uniform_1986}.

The output of a random number generator (RNG) can either be employed directly as a source of randomness or used as input to a pseudorandom number generator (PRNG) to enhance statistical properties or uniformity. When used in its raw form, the output must satisfy stringent randomness criteria, typically assessed through a series of statistical tests. These tests are essential to ensure that the underlying physical process generating the randomness exhibits no detectable structure or bias. For instance, a commonly used physical entropy source, such as electronic noise, may exhibit apparent randomness while in reality containing periodic or structured components—such as waveforms or other regular patterns—that fail to meet statistical thresholds for true randomness. Identifying and filtering such non-random contributions is crucial when RNG outputs are used without further post-processing.

Beyond generation, testing the quality of random numbers is equally critical \cite{lecuyer_history_2017, blass_randomness_2020,knuth_art_1997}. It is impossible to definitively prove a sequence's randomness, as any sequence could theoretically occur. However, Kendall and Babington-Smith \cite{kendall_randomness_1938} proposed the concept of \textit{local randomness}, which states that any reasonably long segment of a random sequence should appear random and pass statistical tests. Statistical tests are used to find evidence against the null hypothesis that the sequence is random.

Calorimeters are fundamental in particle physics experiments, measuring particle energy by absorbing particles and converting their energy into detectable signals. They are categorized as \glspl{ecal}, which measure the energy of electrons and photons through electromagnetic interactions, or \glspl{hcal}, which measure hadronic energy. While the fundamentals of calorimetry are beyond this work's scope, excellent reviews are available \cite{fabjan_calorimetry_2003,leroy_principles_2009}. Recently, the \gls{decal} emerged as a novel readout sensor for calorimetry designed for future particle physics experiments. Unlike traditional calorimeters, the \gls{decal} reconstructs incident energy by counting shower particles instead of measuring deposited energy \cite{allport_first_2020, allport_decal_2022}.

A time series is defined as an ordered sequence of values, typically indexed by time, $z_t$. Time series analysis uses statistical techniques to extract meaningful information from such data, broadly divided into linear \cite{box_time_2016, hamilton_time_1994} and nonlinear \cite{kantz_nonlinear_2003} methods. Linear methods generally assume each data point is a linear combination of previous points and a random shock term, modeled as an independent Gaussian variable with zero mean and constant variance. Nonlinear methods, by contrast, can capture more complex relationships.

This work explores the potential of using noise from the \gls{decal} sensor as a \gls{trng}. Linear time series analysis techniques are employed to extract the stochastic component of the signal, and the extracted noise is tested for quality using statistical methods. This approach has two broader implications: first, the methodology can be extended to extract the stochastic component of noise from similar data sources beyond the \gls{decal} sensor. Second, analyzing the \gls{decal} sensor noise through time series analysis enables its full characterization, up to a stochastic term. This could be valuable for future experiments using the \gls{decal} sensor by enabling more accurate calibration.

\section{Method}
\label{sec:method}

\subsection{Experimental Setup}
\label{subsec:method:experimental-setup}

\begin{figure}
    \centering
    \begin{subfigure}[b]{0.45\textwidth}
        \centering
        \includegraphics[width=\textwidth]{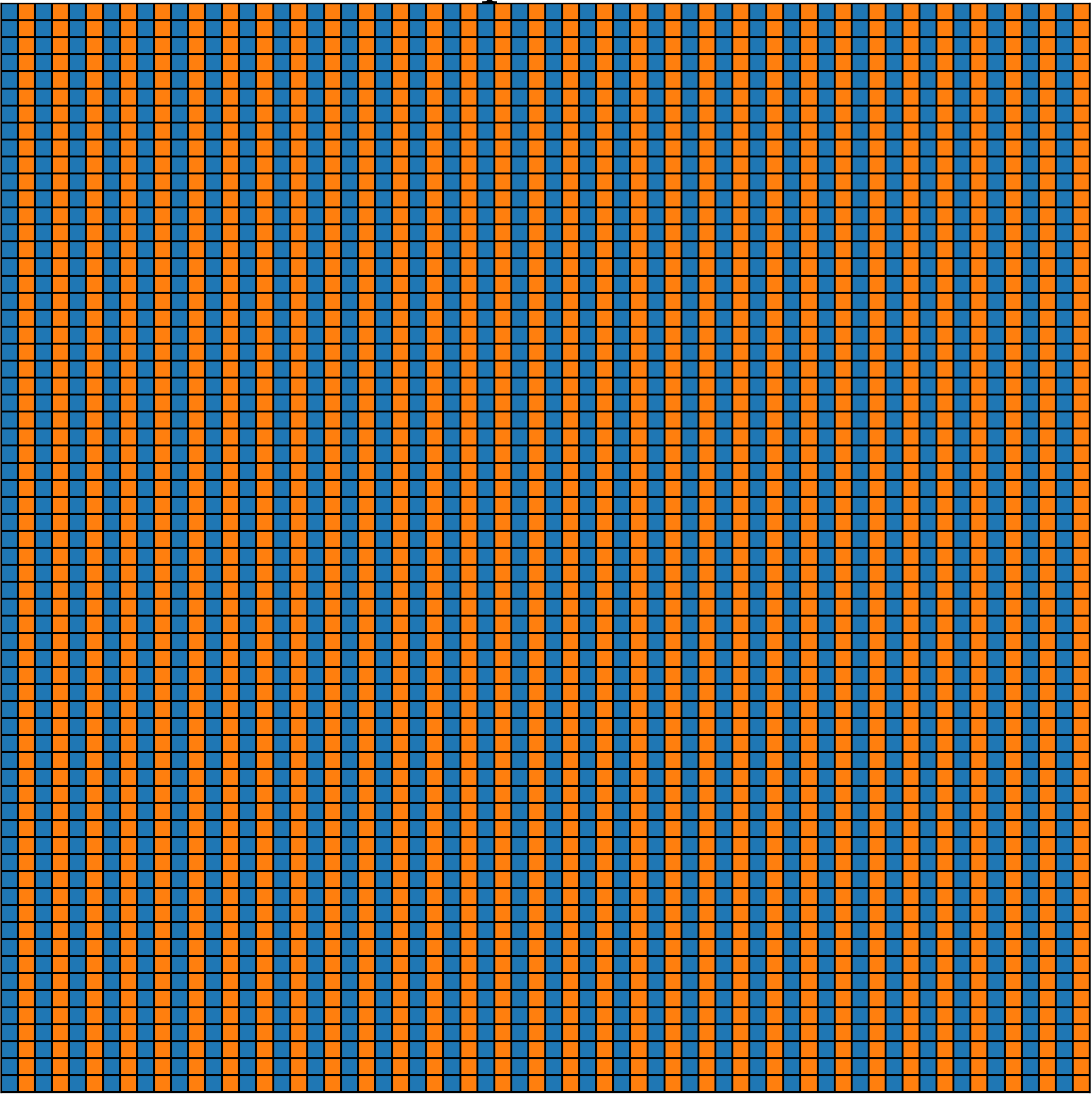}
        \caption{}
        \label{fig:strip-mode}
    \end{subfigure}
    \hfill
    \begin{subfigure}[b]{0.45\textwidth}
        \centering
        \includegraphics[width=\textwidth]{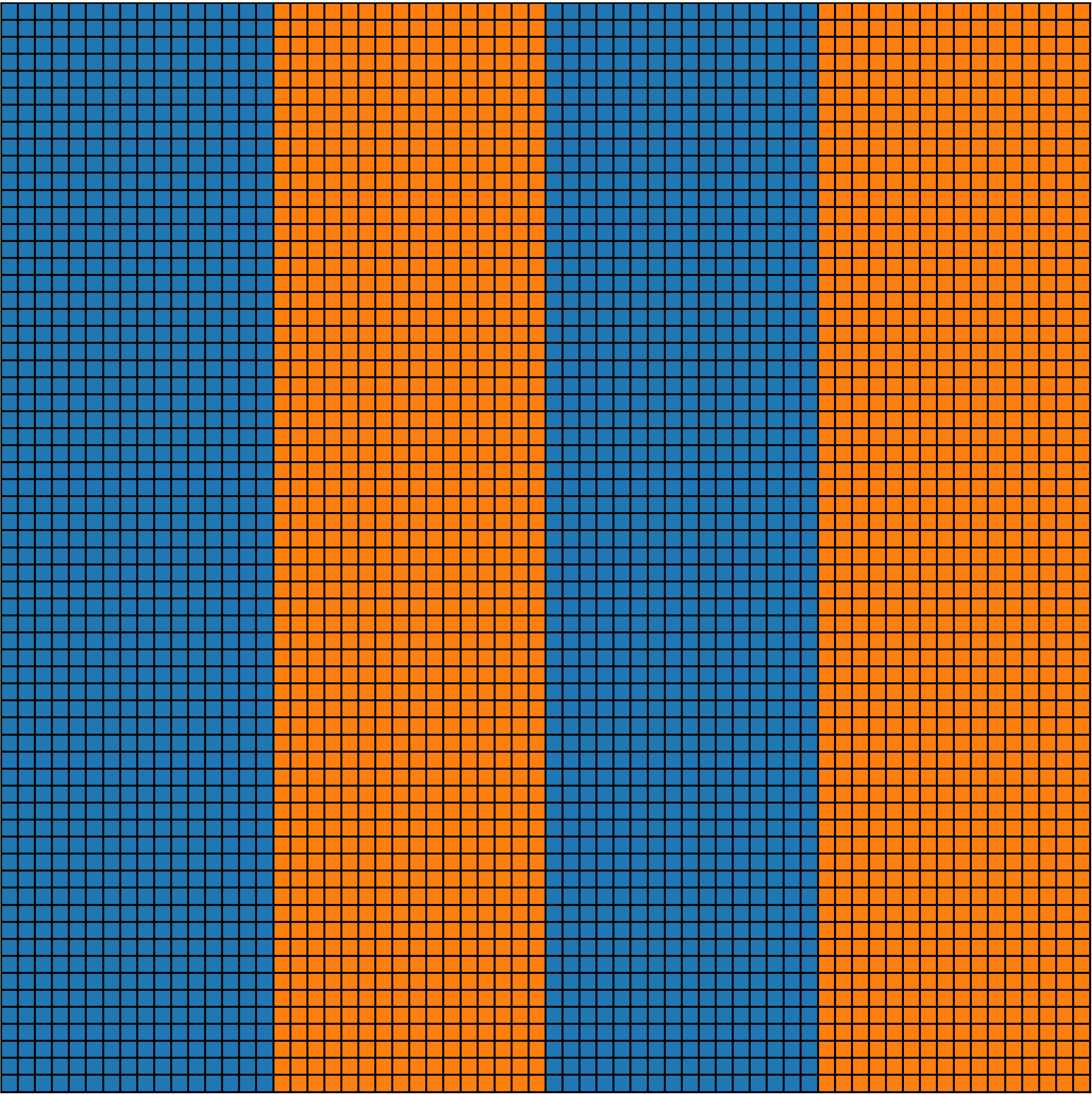}
        \caption{}
        \label{fig:pad-mode}
    \end{subfigure}
    \caption{The \gls{decal} sensor readout modes: (a) strip mode and (b) pad mode.}
    \label{fig:readout-modes}
\end{figure}

The \gls{decal} sensor is a \gls{dmaps} prototype consisting of $64 \times 64$ pixels with $\SI{55}{\micro m}$ pitch and an epitaxial layer of $\SI{25}{\micro m}$ that is expected to be fully depleted when a low bias voltage is applied. Its pixel readout comprises a comparator that detects a hit when the shaper output falls below a globally set threshold voltage. Only this binary information is readout per $\SI{40}{\mega Hz}$ clock cycle. The reconfigurable readout groups the pixel hits in either $64$ strips, each of size $1 \times 64$ pixels (termed \textit{strip mode}), or in four pads of size $16 \times 64$ pixels (termed \textit{pad mode}), as shown in \cref{fig:readout-modes}. In this work, we use the strip mode readout mode, where the chip reads out the pixels in an entire column. The communication between \gls{decal} sensor and computer functions via the Nexys Video \gls{fpga} board and follows the ROOT-based framework of \gls{itsdaq} \cite{atlas_collaboration_technical_2017}. The \gls{decal} sensor is placed in an uncontrolled office desk, that mimics real-life everyday conditions. The technical characteristics of the \gls{decal} sensor are presented in detail in refs. \cite{allport_decal_2022,allport_first_2020,fasselt_characterization_2023,fasselt_energy_2023,kopsalis_evaluation_2022}.

\subsection{Data Acquisition and Preprocessing}
\label{subsec:method:data-acquisition-preprocessing}

The primary data are acquired by \textit{threshold scans}. During these scans, the threshold voltage is varied from a low to a high value, and the comparator of the pixel counts only when the shaper output is near the threshold voltage. Therefore, a distribution of counts of different voltages is being generated. We focus on data obtained from a single pixel of the $64 \times 64$ sensor array, specifically the pixel at column/strip $31$ and row $39$. This choice is arbitrary, as similar results were observed for other pixels as well \cite{aslanis_time_2024}. For this application, sensor tuning (the process of tuning is discussed in refs. \cite{aslanis_time_2024,fasselt_energy_2023,kopsalis_evaluation_2022}) is unnecessary since we are solely interested in the noise and do not aim to filter it. However, to improve efficiency, we tune the pixels to a voltage of approximately $\SI{1.16}{\volt}$, allowing us to scan over a smaller range of threshold voltages. For each threshold scan, we sample the threshold voltage from $\SI{1.10}{\volt}$ to $\SI{1.18}{\volt}$ in steps of $\SI{0.002}{\volt}$, and repeat the scan $2000$ times. Between two consecutive scans, we wait for $200$ clock cycles to ensure the stabilization of the sensor. The data are stored in an ASCII file, and consist of the threshold voltage and the number of counts for each threshold voltage for each scan. We consecutively repeat this process. As noted in previous works \cite{kopsalis_evaluation_2022}, the threshold scans produce a Gaussian-like distribution of counts. We fit a Gaussian function to the distribution using the \gls{bmle} method \cite{barlow_practical_2020} to extract the mean of the distribution. After this process, we obtain a time series of the mean of the Gaussian distribution of the counts. These are the data we use for the time series analysis.

\subsection{Time Series Analysis}
\label{subsec:method:time-series-analysis}

Before proceeding with the time series analysis, we need to define the backward shift operator $B$ as $B z_t = z_{t-1}$, and the difference operator $\nabla$ as $\nabla z_t = z_t - z_{t-1}$. We model the time series of the mean of the Gaussian distribution of counts (termed just \textit{time series} from now on) as an \gls{arima} process:

\begin{equation}
    \phi(B) \nabla^d \tilde{z}_t = \theta(B) \alpha_t,
    \label{eq:arima}
\end{equation}

where $\tilde{z}_t$ is the reduced time series (defined as $\tilde{z}_t = z_t - \mu$ with $\mu$ the mean of the time series). The \gls{arima} process is a combination of an \gls{ar} of order $p$, a \gls{ma} of order $q$, and differencing of order $d$. The \gls{ar} and \gls{ma} operators are defined as $\phi(B) = 1 - \phi_1 B - \phi_2 B^2 - \ldots - \phi_p B^p$ and $\theta(B) = 1 + \theta_1 B + \theta_2 B^2 + \ldots + \theta_q B^q$, respectively, where $\phi_i$ and $\theta_i$ are the coefficients that need to be estimated. The residuals $\alpha_t$ are assumed to be white noise with zero mean and constant variance. Every \gls{arima} process can be fully described by its orders $p$, $d$, and $q$, and the coefficients $\phi_i$ and $\theta_i$. An \gls{arima} process is characterized by two main functions. The first is the \gls{acf}:

\begin{equation}
    \label{eq:autocorrelation_def}
    \rho_{j} \equiv \frac{\gamma_{j}}{\gamma_0},
\end{equation}

where $\gamma_{jt}$ is the autocovariance function:

\begin{equation}
    \label{eq:autocovariance_def}
    \gamma_{jt} \equiv E[(z_t - \mu)(z_{t-j} - \mu)],
\end{equation}

where $E$ is the expectation operator, $E[\cdot] \equiv \lim_{N \to \infty} \frac{1}{N} \sum_{t=1}^{N} (\cdot)$, and $\mu$ is the mean of the time series, $\mu \equiv E[z_t]$. The second function is the \gls{pacf}, which is the correlation between $z_t$ and $z_{t-j}$ after removing the effect of the intermediate values $z_{t-1}, z_{t-2}, \ldots, z_{t-j+1}$, and is rigorously defined in ref. \cite{hamilton_time_1994}. 

The \textit{Box-Jenkins methodology} \cite{box_time_2016} is the most widely used approach for identifying, estimating, and diagnosing \gls{arima} models. It consists of the following steps: (1) difference the time series until it is \textit{stationary} - meaning that the mean and \gls{acf} are constant in time, (2) identify the \gls{ar} and \gls{ma} orders of the stationary time series by examining the \gls{acf} and \gls{pacf}, (3) estimate the parameters of the \gls{arima} model by fitting the model to the data, and (4) perform diagnostic checks on the residuals of the model. The \gls{arima} model is considered valid if the residuals are white noise.

The interplay between the \gls{arima} process and the \gls{acf} and \gls{pacf} which is used in the Box-Jenkins methodology is explored in detail in refs. \cite{box_time_2016,hamilton_time_1994,aslanis_time_2024}. In this work, we use the Box-Jenkins methodology to identify the differencing order $d$, and subsequently use the \gls{aic} \cite{akaike_new_1974} and the \gls{bic} \cite{schwarz_estimating_1978} to identify the \gls{ar} and \gls{ma} orders $p$ and $q$. We then estimate the parameters of the \gls{arima} model using the \textit{innovations Maximum Likelihood Estimation} algorithm, details of which can be found in~\cite{brockwell_introduction_2016}. In order to avoid the effect of any long time nonlinear correlations or seasonality, we regularly re-fit the model to the time series. For the computations, we use the \texttt{statsmodels} library~\cite{seabold_statsmodels_2010} in Python.

\subsection{Random Number Generation}
\label{subsec:method:random-number-generation}

After fitting the \gls{arima} model to the time series, we can extract the white noise component of the model, using Eq. \eqref{eq:arima}:

\begin{equation}
    \label{eq:residuals}
    \hat{\alpha}_{t} = \hat{\theta}^{-1}(B) \hat{\phi}(B) \tilde{w}_{t}
\end{equation}

where $\tilde{w}_{t}$ is the observations after differencing and subtracting the mean. The $\hat{\cdot}$ symbol indicates that these are the estimated values of the parameters.

The white noise component is expected to be a time series of uncorrelated, normally distributed random numbers. In principle, this is the end goal of this work: these are hardware generated true random numbers. There are many ways to generate random binary numbers from these residuals. Since this work focuses on exploring the potential of the \gls{decal} sensor to generate random numbers, a conservative approach is used. As the residuals are normally distributed around zero, the simplest method is to assign a bit to each residual. If the residual is positive, the bit is set to $1$; otherwise, it is set to $0$. This process creates a binary string with a length equal to that of the time series.

This process should produce completely random and uncorrelated bits, provided the time series is an \gls{arima} process and the model fit is accurate. However, to address potential small nonlinearities or deviations between the estimated residuals and the true residuals, an additional shuffle step is implemented. This shuffle relies solely on information from the generated binary string and can be performed continuously and computationally efficiently during random bit generation. The algorithm is as follows:

\begin{figure}[h]
    \centering
    \includegraphics[width=0.5\textwidth]{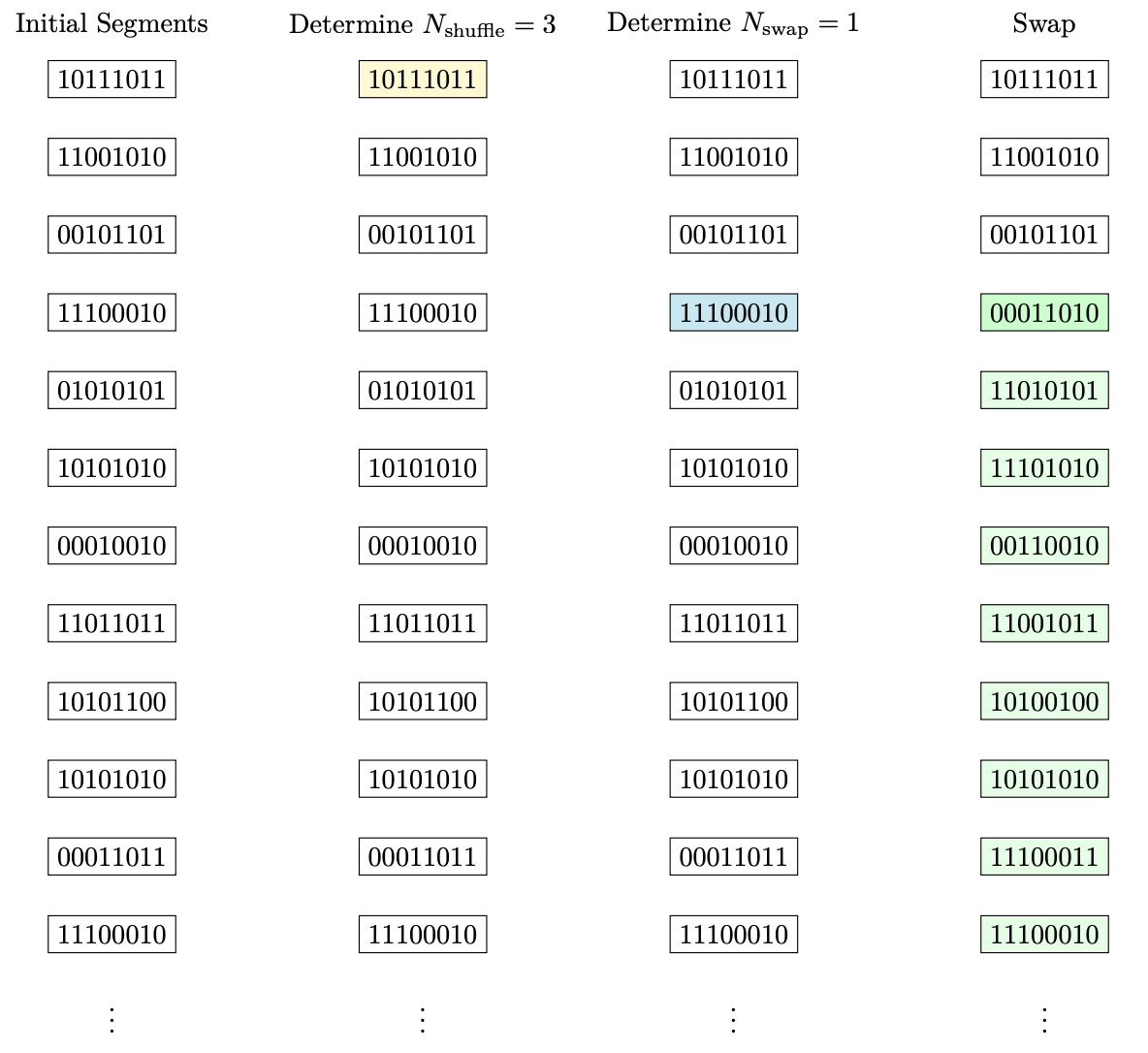}
    \caption{Example for the swap algorithm. See text for details.}
    \label{fig:scheme_swap_algorithm}
\end{figure}

\begin{enumerate}
    \item \textbf{Segment Division}: Divide the binary string into segments of length $L = 1$ byte (i.e., $8$ bits). Iterate over each segment to perform the bit manipulation operations.
    
    \item \textbf{Determine Shuffle Factor}:
        \begin{enumerate}
            \item Identify the first occurrence of the bit `1' within each segment.
            \item Convert the two bits immediately following this `1' into a decimal value, then add 1 to obtain $N_{\mathrm{shuffle}}$, a number between 1 and 4.
        \end{enumerate}
    
    \item \textbf{Skip Segments}: Skip the next $N_{\mathrm{shuffle}}$ segments without any modifications. When the $N_{\mathrm{shuffle}}$-th segment is reached, proceed to the next step.
    
    \item \textbf{Determine Swap Factor}:
        \begin{enumerate}
            \item Within the $N_{\mathrm{shuffle}}$-th segment, locate the first occurrence of the bit `0'.
            \item Convert the two bits immediately following this `0' into a decimal value, then add 1 to obtain $N_{\mathrm{swap}}$, a number between 1 and 4.
        \end{enumerate}
    
    \item \textbf{Bit Swapping}: Swap bits between segments as follows:
        \begin{enumerate}
            \item Swap the first bit of the $N_{\mathrm{shuffle}}$-th segment with the first bit of the $(N_{\mathrm{shuffle}} + N_{\mathrm{swap}})$-th segment.
            \item Swap the second bit of the $N_{\mathrm{shuffle}}$-th segment with the second bit of the $(N_{\mathrm{shuffle}} + 2 \times N_{\mathrm{swap}})$-th segment.
            \item Continue this process for all bits in the segment.
        \end{enumerate}
    
    \item \textbf{Repeat the Process}: Repeat the above steps until the end of the binary string is reached.
\end{enumerate}

\subsection{Random Number Testing}
\label{subsec:method:random-number-testing}

A wide range of statistical tests can be employed to assess whether a given sequence exhibits characteristics consistent with those of an ideal random sequence. Since randomness is fundamentally a probabilistic concept, the expected behavior of a truly random sequence under such tests can be described in statistical terms. Each test is designed to detect specific types of structure or regularity—features whose presence would suggest a deviation from randomness. Given the infinite variety of possible patterns, there exists an unbounded number of statistical tests, and thus no finite collection of tests can be considered exhaustive. Consequently, even sequences that pass many commonly used tests may still harbor undetected regularities. Moreover, the interpretation of test results requires careful statistical reasoning, as misinterpretation may lead to unjustified conclusions regarding the randomness or reliability of a particular generator.

Randomness is typically evaluated by testing a well-defined null hypothesis that assumes the presence of randomness in the sequence under consideration. This is done using carefully constructed statistical tests, each associated with a specific test statistic designed to capture deviations from random behavior. Under the assumption that the null hypothesis holds, the test statistic follows a known theoretical distribution. Based on this reference distribution, a critical threshold is established. The test statistic is then calculated from the observed data and compared to the critical value. If the test statistic exceeds the threshold, the null hypothesis of randomness is rejected, indicating potential structure or non-randomness in the sequence. Conversely, if the test statistic falls within the acceptable range, the null hypothesis is not rejected, and the data are considered to have passed the randomness test.

In order to test the randomness of the  generated numbers from our setup, we first evaluate the residuals and then the binary numbers derived from them. Finally, these binary numbers are tested for randomness using the \gls{nist} Statistical Test Suite \cite{rukhin_statistical_2010}, which includes $15$ tests for randomness.


The residuals of the \gls{arima} model are expected to be normally distributed with a zero mean, constant variance, and no temporal correlation. A visual inspection of the residuals time series and their histogram can provide a first indication of their properties. For a white noise process, the \gls{acf} is expected to be normally distributed around zero, with a standard deviation of $1/\sqrt{n}$, where $n$ is the number of observations \cite{box_time_2016}. Thus, plotting the \gls{acf} of the residuals and verifying that it adheres to these limits is also a useful diagnostic tool.

In practice, however, the true residuals $\alpha_{t}$ are not known; only the estimated residuals $\hat{\alpha}_{t}$ are available. It has been shown \cite{box_distribution_1970} that the estimation errors of model parameters can significantly affect the standard errors of the residuals' \gls{acf}, especially for small lag values. Box and Pierce \cite{box_distribution_1970} provided a method to correct the \gls{acf} standard errors for parameter estimation effects. In this work, we opt to neglect this correction.

In addition to the visual inspection of the residuals, Box and Pierce \cite{box_distribution_1970} introduced a test, later modified by Ljung and Box \cite{ljung_measure_1978}, called the \textit{Ljung-Box} test. The test statistic is given by:

\begin{equation}
    \label{eq:ljung_box_pierce}
    Q = n(n+2) \sum_{k=1}^{K} \frac{\hat{\rho}^{2}_{k}}{n-k}.
\end{equation}
    
Here, $n = N - d$ is the number of observations, with $N$ being the total number of observations and $d$ the differencing order of the \gls{arima} model. The parameter $K$ represents the number of lags (set to $K=40$ in this work), $\hat{\rho}_{k}$ is the sample \gls{acf} at lag $k$, and $Q$ is the test statistic. Under the null hypothesis that the residuals are white noise, the test statistic $Q$ follows a $\chi^2$ distribution with $K - p - q$ degrees of freedom, where $p$ and $q$ are the autoregressive and moving average orders, respectively.

Another diagnostic performed is the \textit{cumulative periodogram} test, which originates from \textit{spectral analysis} and is used to identify periodicity in time series data. For a white noise process, the cumulative periodogram should approximate a straight line connecting the points $(0, 0)$ and $(0.5, 1)$. The standard error bounds for the cumulative periodogram are parallel lines at distances $\pm K_{\epsilon}/q$ from the periodogram, where $K_{\epsilon} = 1.36$ for a $95\%$ confidence interval. The parameter $q$ is computed as $(n-2)/2$ for even $n$ and $(n-1)/2$ for odd $n$. Further details on this test can be found in \cite{box_time_2016}.

Next, we test the binary numbers that are generated from the residuals. For an initial check, that is easy to visualize, we simulate a diffusion process, modeled as a random walk. Two versions of the diffusion process are considered. In the one-dimensional diffusion, the walker starts at position $0$ and moves up if the random bit is $1$ or down if the random bit is $0$, with its position recorded at each step. In the two-dimensional diffusion, the walker starts at position $(0,0)$ and moves up, down, left, or right based on pairs of random bits: $00$, $10$, $11$, and $01$, respectively. The walker's position is recorded at each step.

The diffusion process is quantified using the \gls{msd}, defined as:
\begin{equation}
    \label{eq:msd}
    \text{MSD}(t) = \langle (\boldsymbol{x}(t) - \boldsymbol{x}(0))^{2} \rangle \approx \frac{1}{N} \sum_{i=1}^{N} (\boldsymbol{x}_{i}(t) - \boldsymbol{x}_{i}(0))^{2},
\end{equation}

where $\boldsymbol{x}(t)$ is the position of the walker at time $t$, $\boldsymbol{x}(0)$ is the initial position, $N$ is the number of particles, and $\langle \cdot \rangle$ denotes averaging over all walkers. The \gls{msd} is expected to follow a linear relationship with time, with its slope corresponding to the diffusion coefficient.

To evaluate the random walk, the random bits are divided into segments, and the diffusion process is simulated for each segment. Statistics are then gathered and compared to theoretical expectations. The \gls{msd} of the one-dimensional diffusion process is expected to scale linearly with time, with a slope of $1$. An efficient Fast Fourier Transform-based algorithm \cite{calandrini_nmoldyn_2011} is used to compute the \gls{msd}, and pseudo-random Monte Carlo simulations generate $66\%$ and $95\%$ confidence intervals, testing the balance of $1$s and $0$s in the random bit segments.

For the two-dimensional case, the walker independently performs one-dimensional random walks in the $x$ and $y$ directions. Over many steps, the walker's position should be symmetrically distributed around the origin, and its distance from the origin should asymptotically follow a Rayleigh distribution with a scale parameter of $\sqrt{t}$ \cite{rayleigh_problem_1905}. This test evaluates the balance of $00$, $01$, $10$, and $11$ bit pairs in the random segments.

\section{Results}
\label{sec:results}

\subsection{Data Preprocessing}
\label{subsec:results:data-preprocessing}

\begin{figure}[h!]
    \centering
    \begin{subfigure}[b]{0.38\textwidth}
        \centering
        \includegraphics[width=\textwidth]{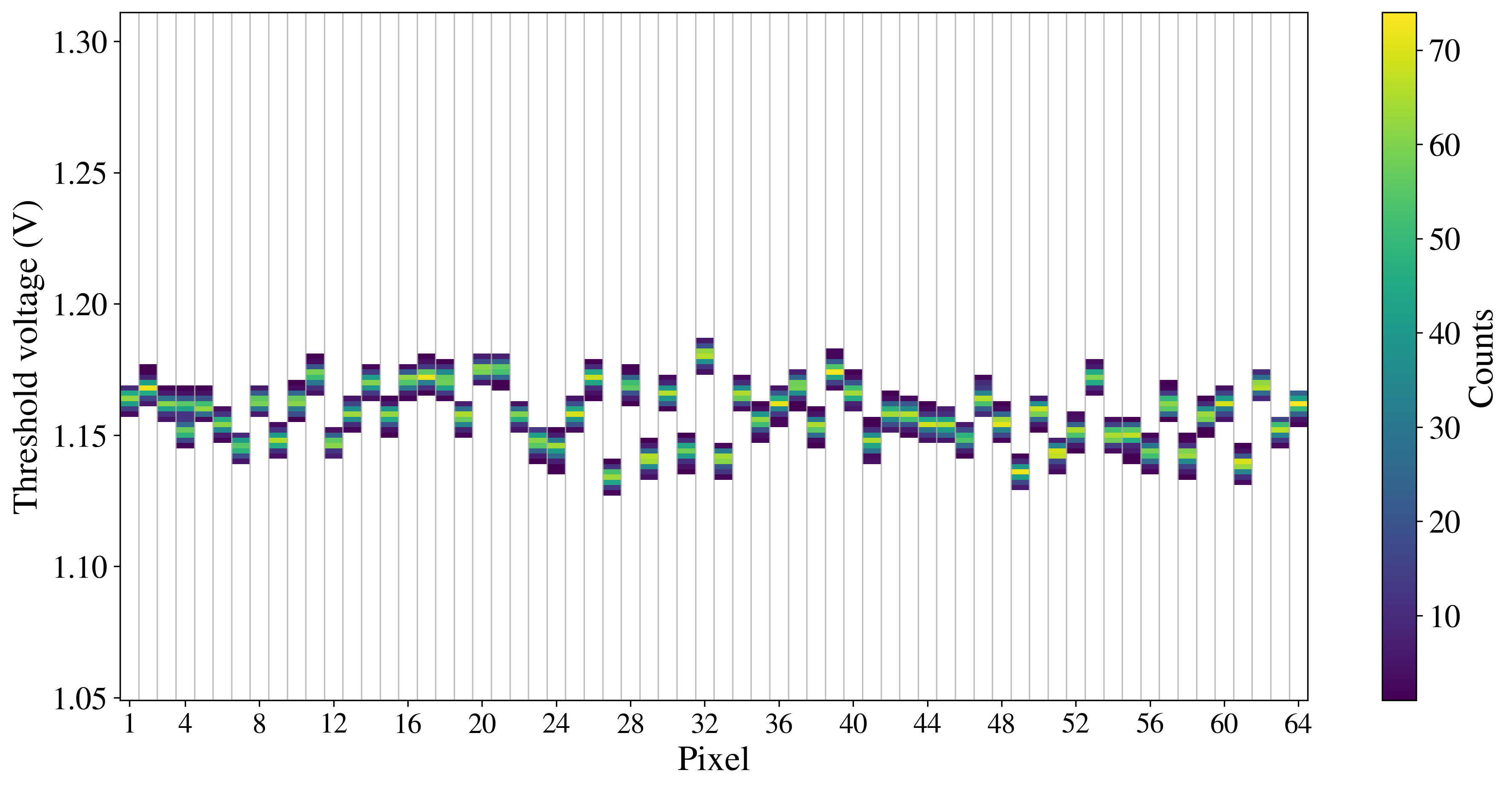}
        \caption{}
        \label{fig:untuned-row}
    \end{subfigure}
    \hfill
    \begin{subfigure}[b]{0.38\textwidth}
        \centering
        \includegraphics[width=\textwidth]{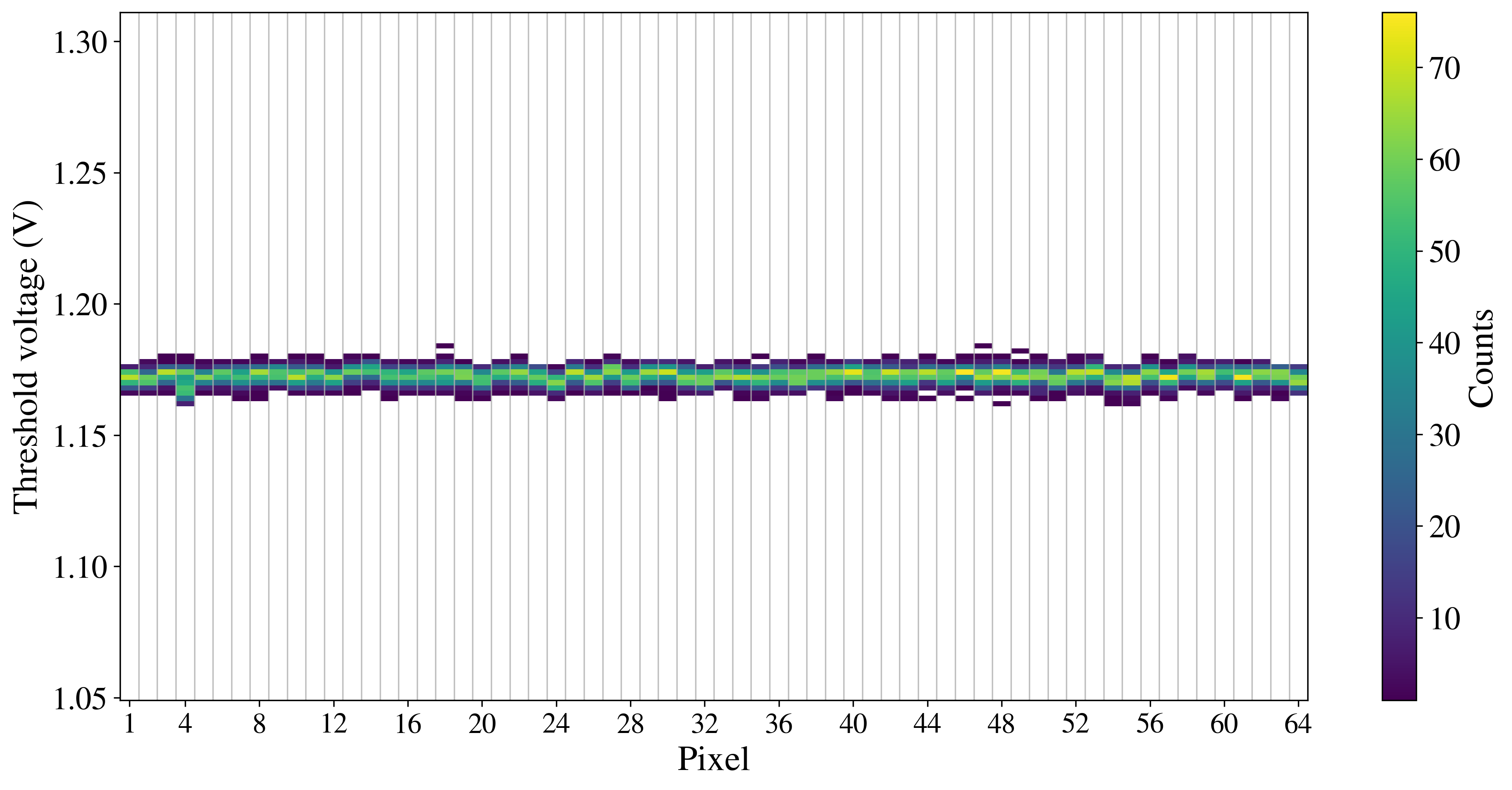}
        \caption{}
        \label{fig:tuned-row}
    \end{subfigure}
    \caption{Histogram of a row of pixels before (a) and after (b) tuning.}
    \label{fig:results:tuned-untuned-row}
\end{figure}

We confirm the observation of previous studies \cite{kopsalis_evaluation_2022} that the threshold scans produce Gaussian distributions, and we generate the time series of their mean values. The outcome of the tuning procedure is depicted in \cref{fig:results:tuned-untuned-row}, which compares the histograms of the pixels of a row before tuning with the corresponding histograms after tuning. The tuning process considerably reduces the range of threshold scan values, thereby enhancing the efficiency of data acquisition by eliminating the need to scan a wide voltage range.

\begin{figure}[h!]
    \centering
    \includegraphics[width=0.6\textwidth]{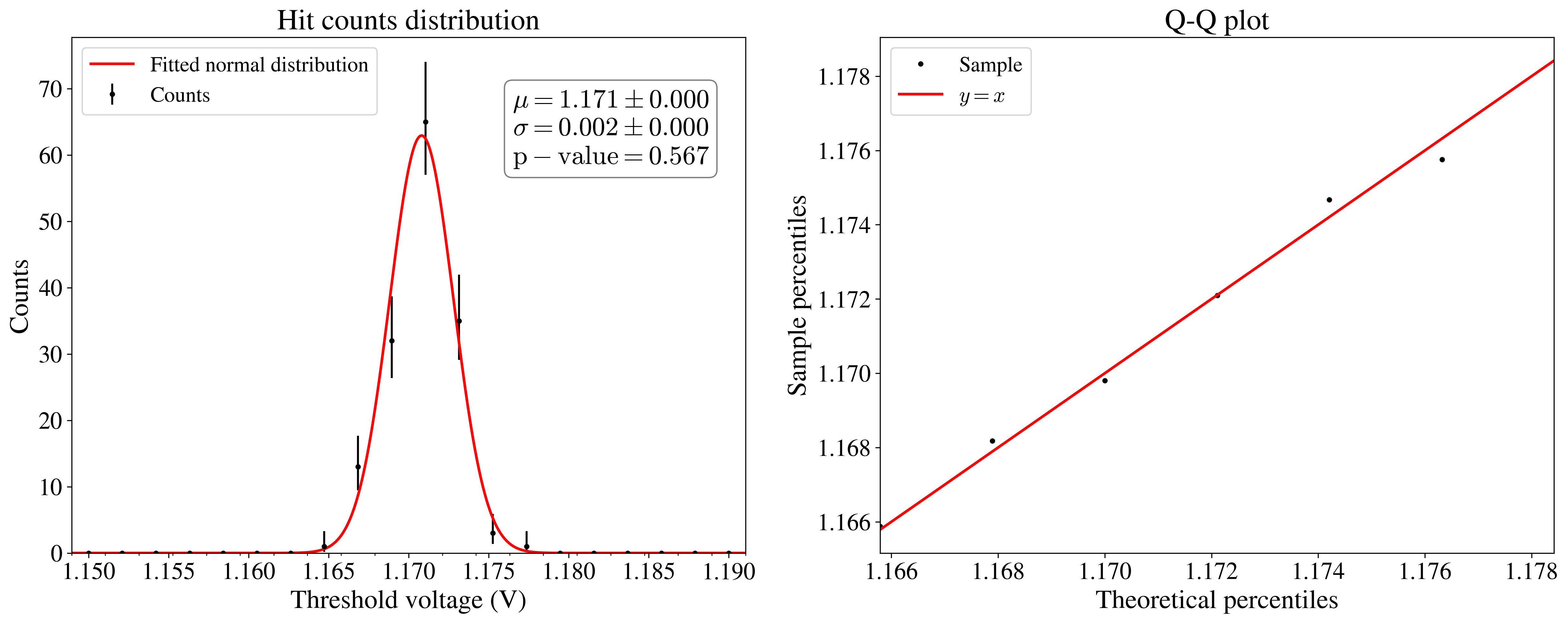}
    \caption{The distribution of counts for a threshold scan (left) and the Q-Q plot that compares the empirical distribution of counts to a Gaussian distribution (right).}
    \label{fig:results:threshold-scan-distribution}
\end{figure}

\subsection{Random Number Generation}
\label{subsec:results:random-number-generation}

\begin{figure}[h!]
    \centering
    \includegraphics[width=0.4\textwidth]{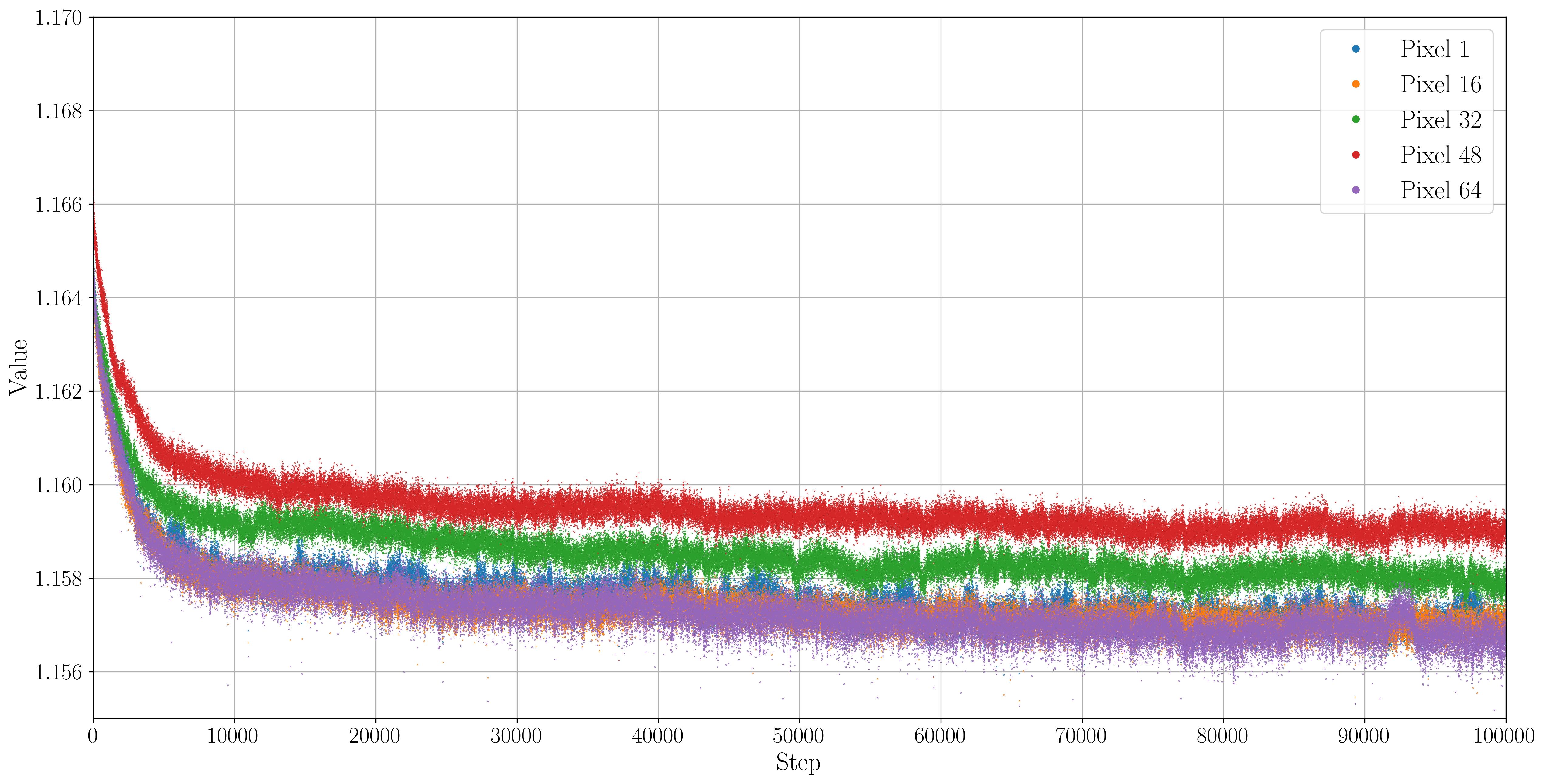}
    \caption{Time series for different pixels of the same row.}
    \label{fig:results:pixel-time-series}
\end{figure}

The time series for different pixels of the same row can be observed in \cref{fig:results:pixel-time-series}. From this figure, it is evident that the time series quickly drops during the first few scans, after which it stabilizes. This is likely due to the warm-up of the sensor, and has also been noted and discussed in previous studies \cite{fasselt_characterization_2023}. In our case, it seems that the sensor stabilizes after around 10000 scans. We therefore discard the first 10000 scans in the rest of the analysis. It can also be seen that the behavior of the time series of the different pixels is similar, as is evident from certain common peaks and valleys in \cref{fig:results:pixel-time-series}. The same observation can be made from similar plots in the literature \cite{fasselt_characterization_2023}. The correlation between the time series of different pixels is further explored in ref. \cite{aslanis_time_2024}.

\begin{figure}[h!]
    \centering
    \includegraphics[width=0.8\textwidth]{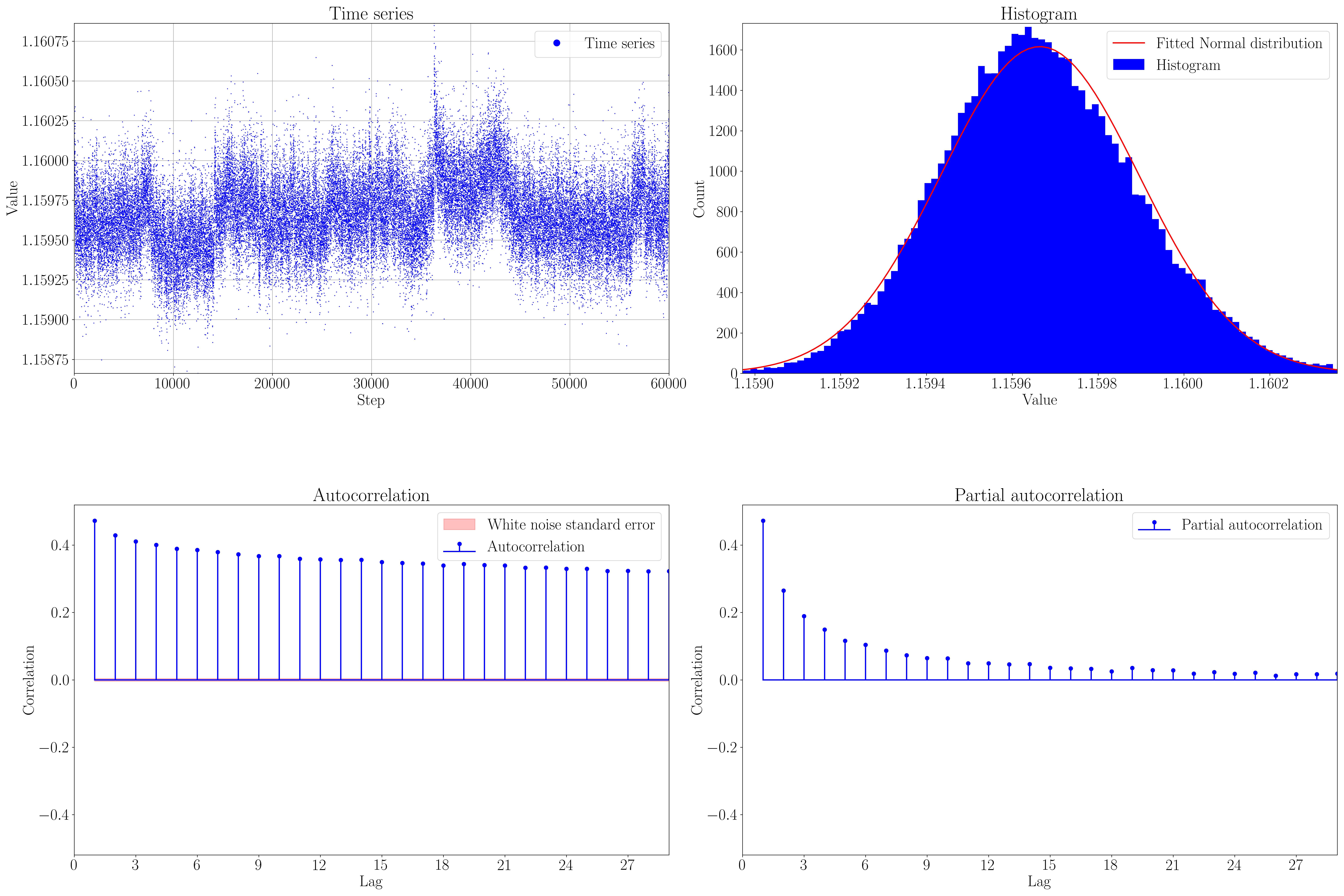}
    \caption{Example of time series (top left), histogram (top right), \gls{acf} (bottom left), and \gls{pacf} (bottom right) for a pixel of the \gls{decal} sensor after discarding the first $10000$ scans. The white noise standard error of the \gls{acf} (in which $66.6\%$ of the values should lie) is calculated as $1/\sqrt{N}$, where $N$ is the number of samples \cite{box_time_2016}.}
    \label{fig:results:raw-time-series}
\end{figure}

\begin{figure}[h!]
    \centering
    \includegraphics[width=0.8\textwidth]{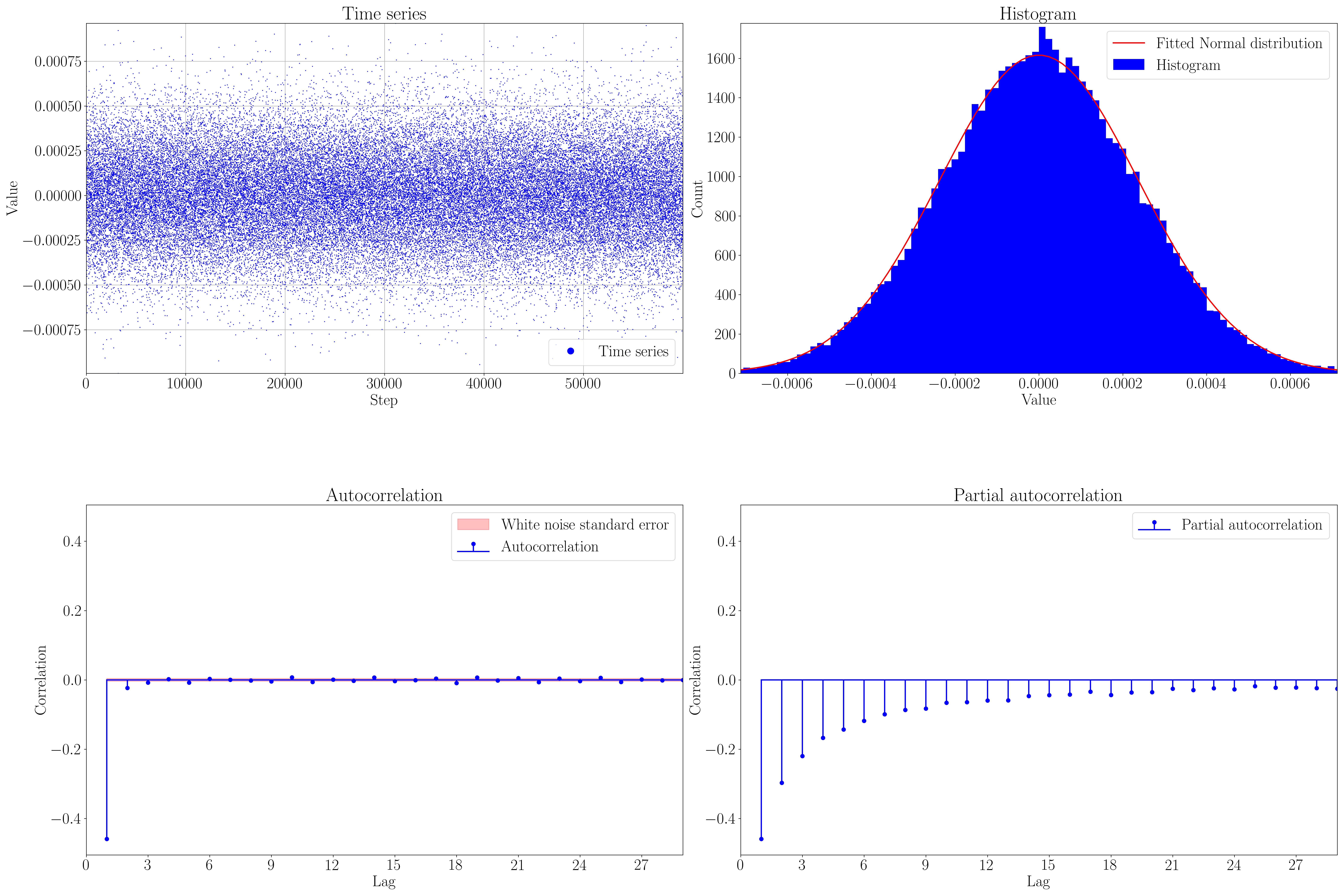}
    \caption{Example of time series (top left), histogram (top right), \gls{acf} (bottom left), and \gls{pacf} (bottom right) for a pixel of the \gls{decal} sensor after discarding the first $10000$ scans and differencing the time series once. The white noise standard error of the \gls{acf} is calculated as in \cref{fig:results:raw-time-series}.}
    \label{fig:results:differenced-time-series}
\end{figure}

\begin{figure}[h!]
    \centering
    \includegraphics[width=0.8\textwidth]{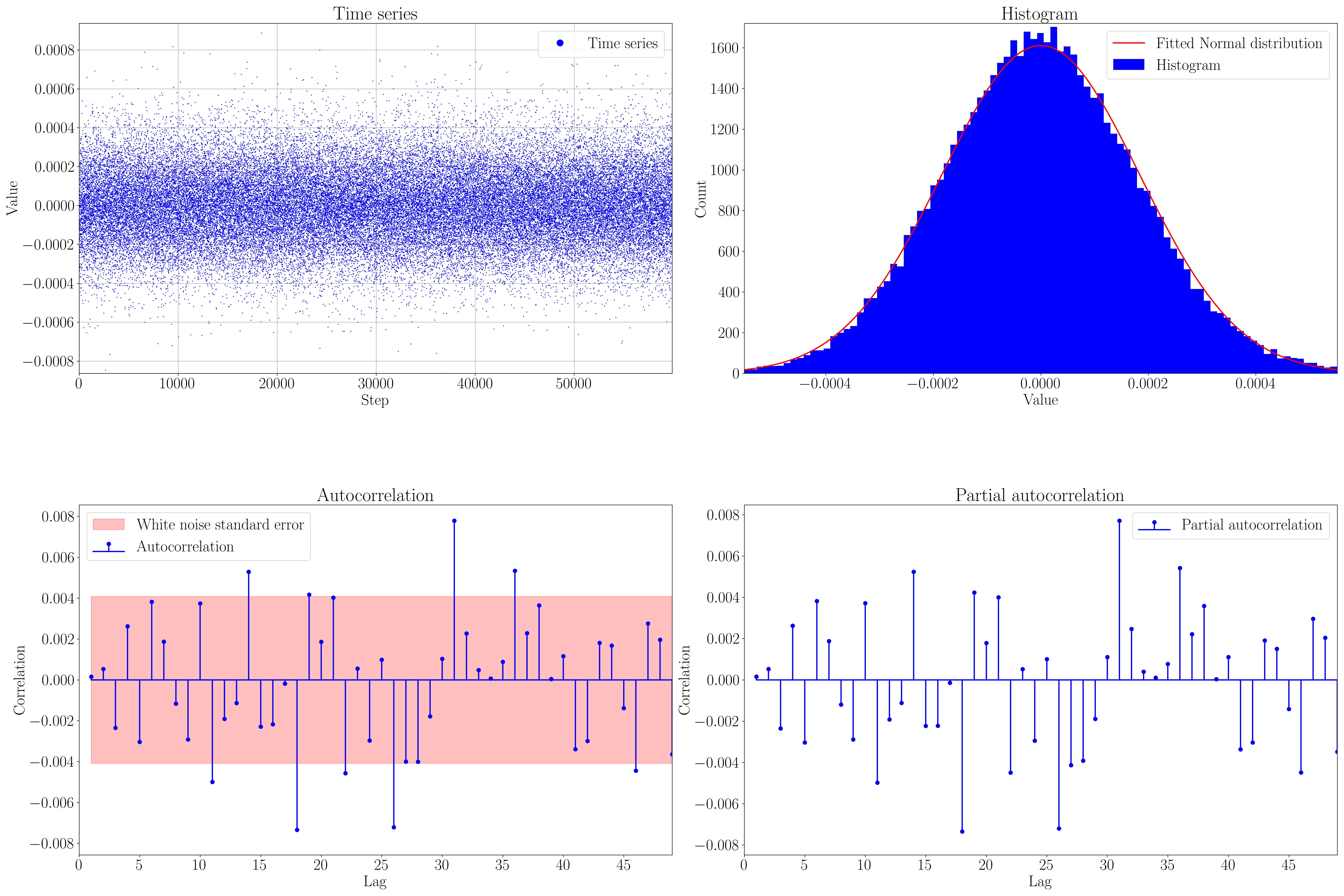}
    \caption{Example of time series (top left), histogram (top right), \gls{acf} (bottom left), and \gls{pacf} (bottom right) for a pixel of the \gls{decal} sensor after discarding the first $10000$ scans and fitting an \gls{arima}$(3,1,5)$ model. The white noise standard error of the \gls{acf} is calculated as in \cref{fig:results:raw-time-series}.}
    \label{fig:results:residuals-time-series}
\end{figure}

Following the Box-Jenkins methodology described in \cref{sec:method}, we calculate the \gls{acf} and \gls{pacf} of the time series, which are shown along with its histogram in \cref{fig:results:raw-time-series}. The slow and linear decrease of the \gls{acf} and \gls{pacf} indicates that the time series is non-stationary. According to the Box-Jenkins methodology, we can make the time series stationary by differencing. The differenced time series along with its \gls{acf}, \gls{pacf} and histogram is shown in \cref{fig:results:differenced-time-series}. By combining the \gls{aic} and \gls{bic} criteria to identify candidate orders and subsequently verifying the residuals for white noise, we determine that the optimal order for the \gls{ar} and \gls{ma} terms is $3$ and $5$, respectively. The residuals of the fitted \gls{arima} model are shown in \cref{fig:results:residuals-time-series}.

\subsection{Validation of Randomness}
\label{subsec:results:validation-randomness}

\begin{figure}[h!]
    \begin{subfigure}[b]{0.35\textwidth}
        \includegraphics[width=\textwidth]{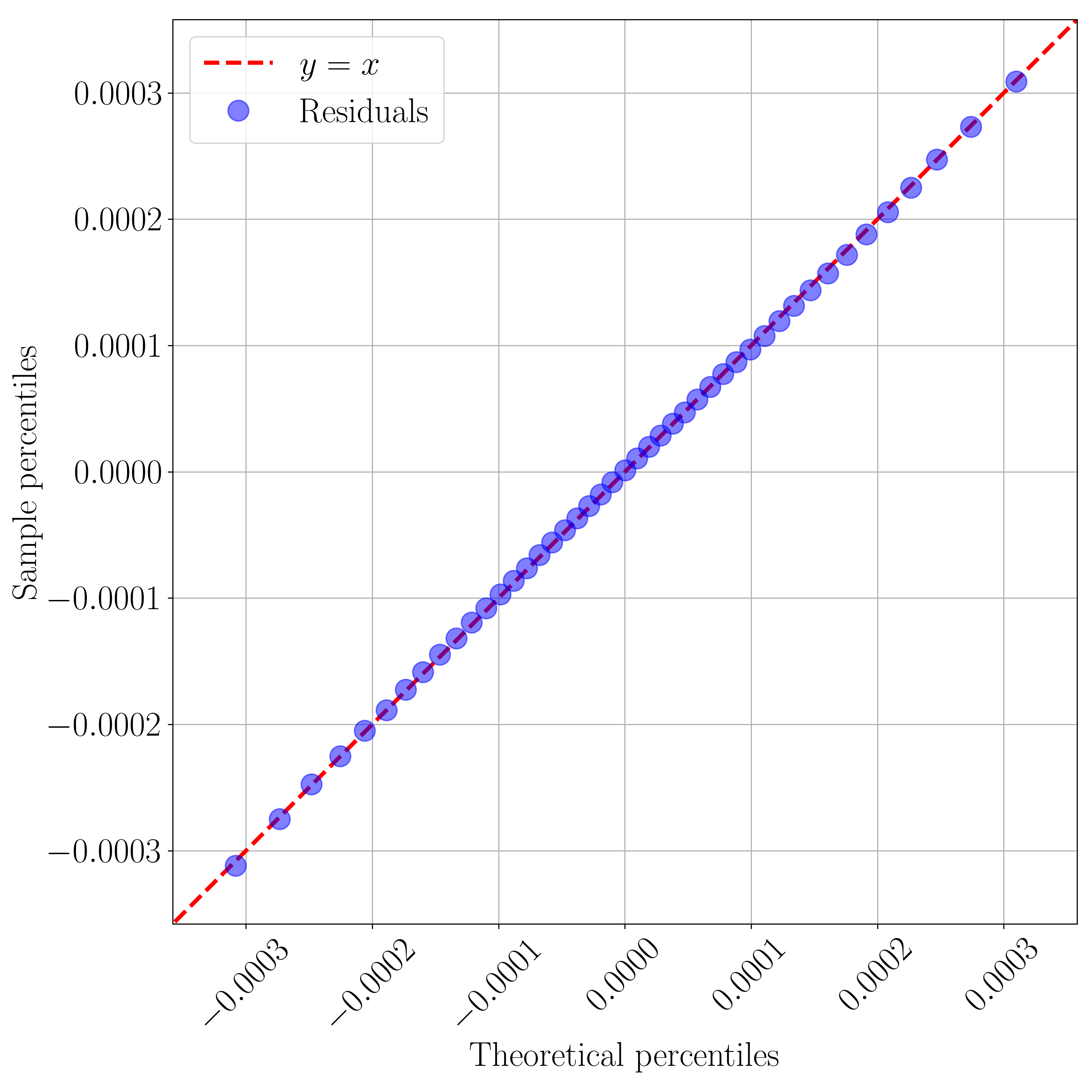}
        \label{fig:results:qq-plot}
    \end{subfigure}
    \hfill
    \begin{subfigure}[b]{0.35\textwidth}
        \includegraphics[width=\textwidth]{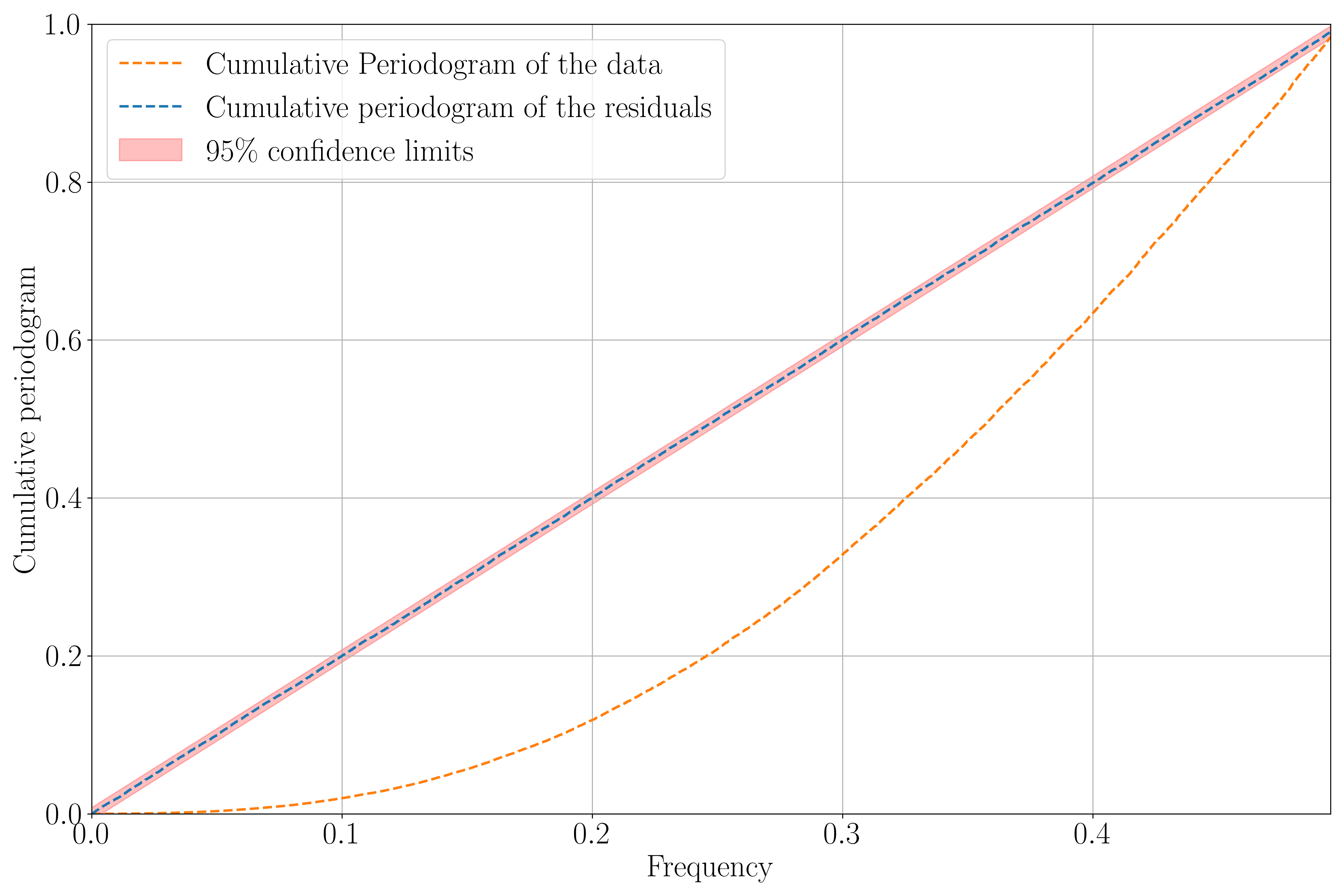}
        \label{fig:results:periodogram}
    \end{subfigure}
    \caption{Example of a (a) Q-Q plot and (b) cumulative periodogram for the residuals of the \gls{arima} model fitted to the time series of a pixel of the \gls{decal} sensor.}
    \label{fig:results:qq-plot-periodogram}
\end{figure}

\begin{figure}[h!]
    \centering
    \includegraphics[width=0.6\textwidth]{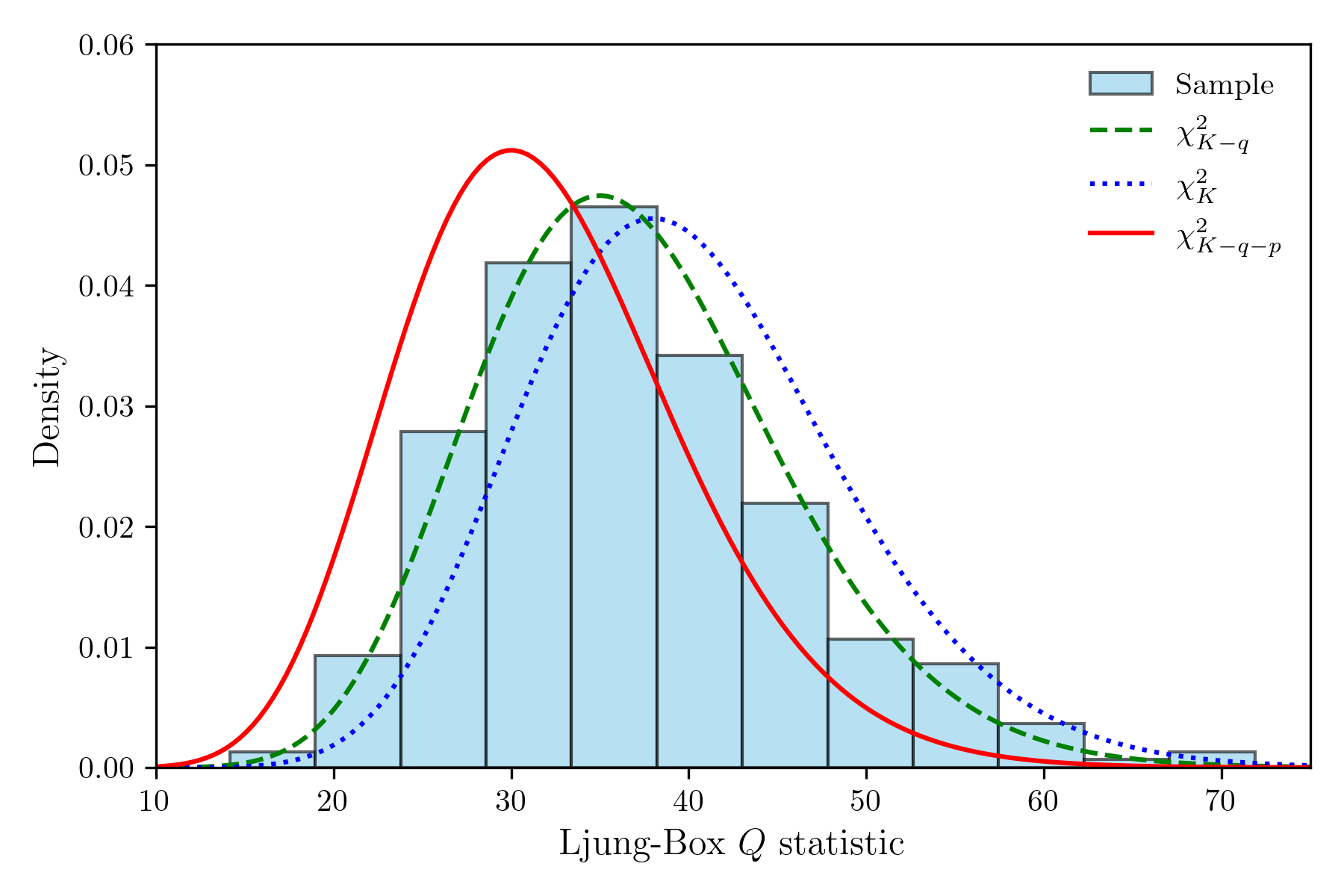}
    \caption{Histogram of the Ljung-Box $Q$ statistic for the residuals of the \gls{arima} model fitted to the time series of a pixel of the \gls{decal} sensor. The red continuous line indicates the $\chi^2$ distribution with $K-q-p$ degrees of freedom, the green dashed line indicates the $\chi^2$ distribution with $K-q$ degrees of freedom, and the blue dotted line indicates the $\chi^2$ distribution with $K$ degrees of freedom.}
    \label{fig:results:ljung-box}
\end{figure}

The normality of the residuals is evident from \cref{fig:results:residuals-time-series} and further confirmed by the Q-Q plot in \cref{fig:results:qq-plot-periodogram}. The distribution of the Ljung-Box $Q$ statistic for the residuals of the \gls{arima} model is shown in \cref{fig:results:ljung-box}. It is apparent that the $Q$ statistic closely follows a $\chi^2$ distribution with $K-q$ degrees of freedom, rather than the expected $\chi^2$ distribution with $K-q-p$ degrees of freedom. This discrepancy is likely due to a suboptimal choice of the \gls{arima} model order, or the presence of periodic and non-linear contributions in the residuals. However, deviations between the expected and observed degrees of freedom in the $\chi^2$ distribution have been noted by many authors \cite{box_time_2016,danioko_novel_2022,francq_diagnostic_2005,davies_significance_1977}. Addressing this discrepancy is outside the scope of this work. Given that the $Q$ statistic follows a $\chi^2$ distribution and, as discussed below, the random number generation passes all other statistical tests, we conclude that the residuals of the \gls{arima} model are white noise to the degree of accuracy required for our analysis. It is worth noting, however, that addressing this discrepancy would be important for a more detailed analysis beyond the scope of the feasibility study presented here. Lastly, the absence of periodic contributions in the residuals of the \gls{arima} model is evident from the periodogram shown in \cref{fig:results:qq-plot-periodogram}.

\begin{figure}[h!]
    \centering
    \begin{subfigure}[b]{0.8\textwidth}
        \includegraphics[width=\textwidth]{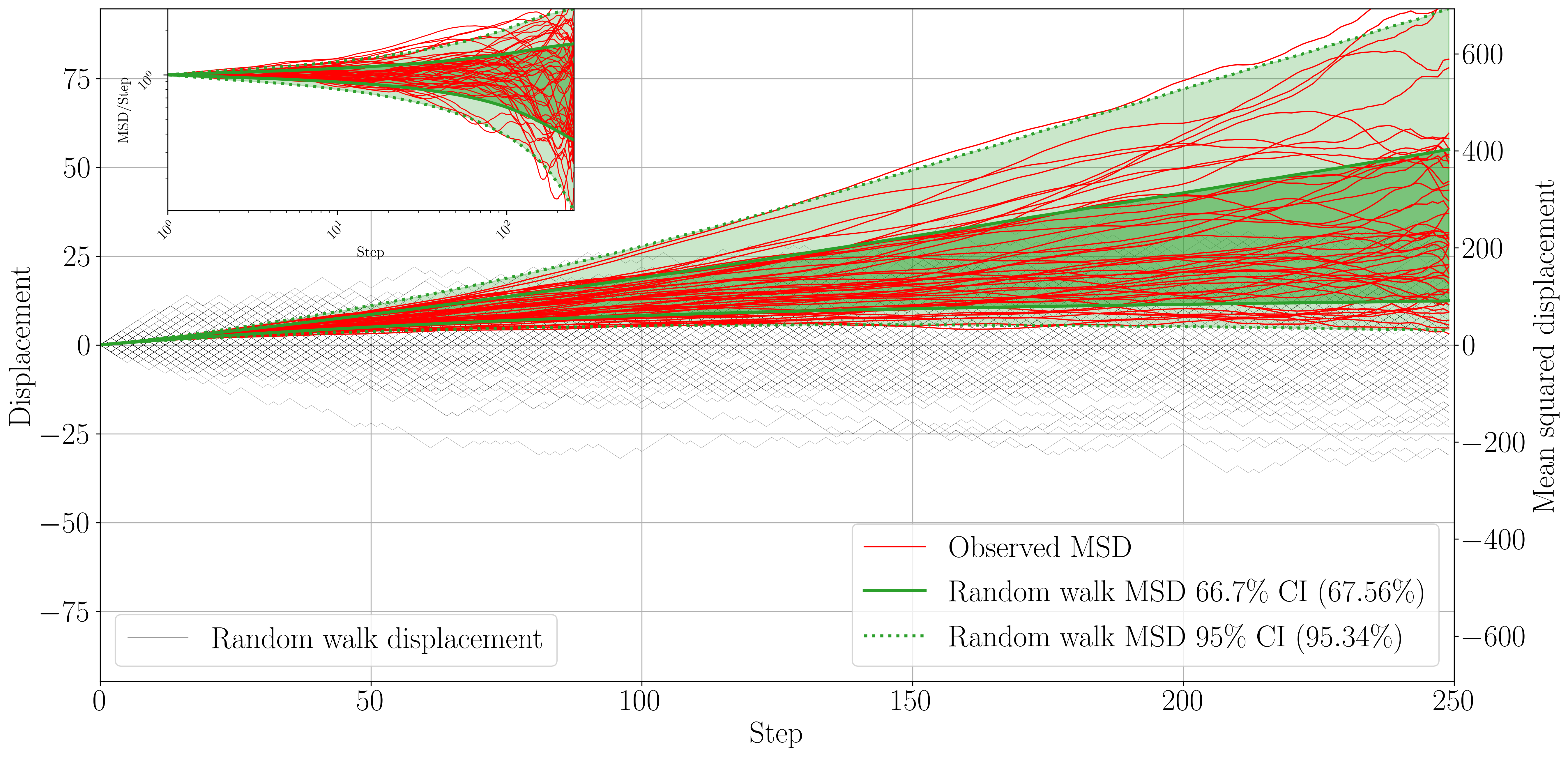}
        \label{fig:results:random-walk}
    \end{subfigure}
    \vfill
    \begin{subfigure}[b]{0.5\textwidth}
        \includegraphics[width=\textwidth]{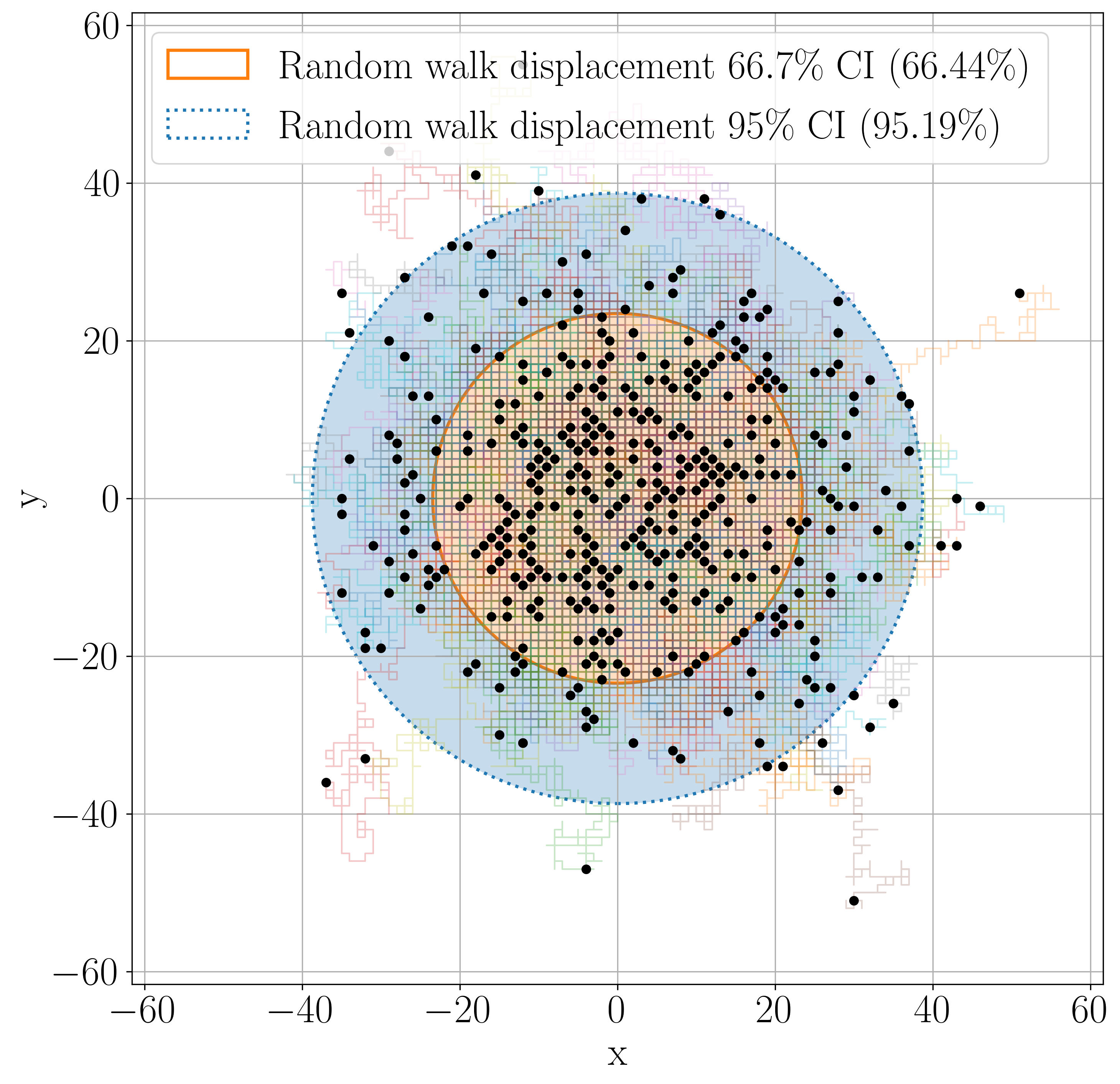}
        \label{fig:results:random-walk-2d}
    \end{subfigure}
    \caption{Results of the (a) one-dimensional and (b) two-dimensional random walk tests. See text for details.}
    \label{fig:results:random-walk-tests}
\end{figure}

The results of the diffusion process tests for the binary data are shown in \cref{fig:results:random-walk-tests}. In both cases, the simulated random walk closely matches the expected behavior. The results of the \gls{nist} tests are summarized in \cref{tab:nist_tests}. For the symbol convention, we follow the one used in the original documentation \cite{rukhin_statistical_2010}: $n$ represents the number of bits in the binary data assessed in each repetition, $m$ or $M$ refers to the relevant parameter for the test, the $p$-value indicates the probability that the underlying $p$-values follow the uniform distribution (with a critical value of $0.0001$), and the number of passes refers to how many times the $p$-value exceeds $0.01$ across repetitions. Some tests have sample size limitations, which is why results are either unavailable or restricted for certain tests. It is evident that the random numbers pass all applicable tests.

\begin{table}
    \centering
    \small
    \begin{tabular}{|c|c|c|c|c|}
        \hline
        Test & n & m/M & p-value & Passes (Minimum) \\
        \hline
        Frequency & $10000$ & - & $0.148094$ & $298/300 \,(291)$ \\
        Block Frequency & $10000$ & $10$ & $0.159613$ & $296/300 \,(291)$ \\
        Runs & $10000$ & - & $0.856907$ & $297/300 \,(291)$ \\
        Longest Run & $10000$ & - & $0.217094$ & $298/300 \,(291)$ \\
        Rank & $38912$ & - & $0.205375$ & $76/77 \,(73)$ \\
        Fourier & $10000$ & - & $0.001801$ & $295/300 \,(291)$ \\
        Non-overlapping & $10^6$ & $9$ & - & $433/441 \,(427)$ \\
        Overlapping & $10^6$ & $9$ & - & $3/3 \,(2)$ \\
        Universal & $10^6$ & - & - & $3/3 \,(2)$ \\
        Linear Complexity & $10^6$ & $500$ & - & $3/3 \,(2)$ \\
        Serial & $10000$ & $2$ & $0.699313, \, 0.127148$ & $292, \, 296/300 \,(291)$ \\
        Approx. Entropy & $10000$ & $2$ & $0.487885$ & $297/300 \, (291)$ \\
        Cumulative Sums & $10000$ & - & $0.514124, \, 0.209577$ & $299, \, 298/300 \,(291)$ \\
        \hline
    \end{tabular}
    \hspace{3em}
    \caption{Results of the \gls{nist} tests. See text for details on the symbol convention. For information on the tests and their limitations, see \cite{rukhin_statistical_2010}.}
    \label{tab:nist_tests}
\end{table}

\subsection{Performance}
\label{subsec:results:performance}

Aside from the quality of the random bits, the efficiency of a \gls{trng}, measured by the rate of high-quality bits generated, is also important. Existing \glspl{trng} can generate random bits at rates exceeding $\SI{10}{Mbps}$~\cite{koc_true_2014,noauthor_quantis_nodate,bucci_high-speed_2003,kim_massively_2021,xu_high_2016}. In the current \gls{decal} setup, the data acquisition process is the slowest step, taking approximately $\SI{310}{min}$ for $100000$ scans. This results in a rate of $\SI{5.4}{bps}$, which is considerably slower than commercial \glspl{trng}. However, the current setup is not optimized for speed; it was designed to test the \gls{decal} as a calorimeter, rather than for random bit generation. It is likely that the rate could be significantly improved by optimizing the data acquisition process, either by redesigning the setup to better suit a \gls{trng}, or by improving the firmware of the sensor. In this work, the focus has been on assessing the quality of the random bits and performing a feasibility study of the \gls{decal} sensor as a \gls{trng}. Optimization of the rate is left for future work, with some suggestions for improvement discussed in \cite{aslanis_time_2024}.

\section{Conclusions}
\label{sec:conclusions}

The feasibility of using the \gls{decal} sensor as a \gls{trng} was evaluated through time series analysis. An \gls{arima} model was found to be an effective fit for the sensor noise. Visual inspection of the residuals, along with diagnostic tests, confirmed the adequacy of the model. The residuals of the time series were then used to generate binary sequences, which were tested for randomness using a diffusion process test and the \gls{nist} suite for randomness testing. The binary sequences passed all tests, indicating that the sensor noise can indeed be used to generate truly random numbers.

Despite the successful generation of random numbers, the process was found to exhibit a low rate of random bit generation, which could limit the sensor's practical application as a \gls{trng}. However, it was noted that the current setup was significantly suboptimal, designed primarily for testing the \gls{decal} sensor as a calorimeter rather than for efficient random number generation. As such, the data acquisition process could be improved to increase the rate of random bit generation. Future work should focus on optimizing the data acquisition process to make the \gls{decal} sensor more suitable for high-speed random number generation.


\printbibliography

\end{document}